# Massively parallel computing on an organic molecular layer

*Anirban Bandyopadhyay[1]\*, Ranjit Pati[2], Satyajit Sahu[1], Ferdinand Peper[3], Daisuke Fujita[1]*

[1]Advanced Nano Characterization Center, National Institute for Materials Science, 1-2-1 Sengen, Tsukuba, Ibaraki, 305-0037 Japan. [2]Department of Physics, Michigan Technological University, Houghton, Michigan, 49931 USA. [3]National Institute of Information and Communications Technology, Kobe, Japan 651-2492.

Current computers operate at enormous speeds of $\sim 10^{13}$ bits/s, but their principle of sequential logic operation has remained unchanged since the 1950s. Though our brain is much slower on a per-neuron base ($\sim 10^3$ firings/s), it is capable of remarkable decision-making based on the collective operations of millions of neurons at a time in ever-evolving neural circuitry. Here we use molecular switches to build an assembly where each molecule communicates–like neurons–with many neighbors simultaneously. The assembly's ability to reconfigure itself spontaneously for a new problem allows us to realize conventional computing constructs like logic gates and Voronoi decompositions, as well as to reproduce two natural phenomena: heat diffusion and the mutation of normal cells to cancer cells. This is a shift from the current static computing paradigm of serial bit-processing to a regime in which a large number of bits are processed in parallel in dynamically changing hardware.

In current computers the logic can reconfigure itself for a new problem or even select a suitable circuit from a few available ones to evolve its hardware.[1] However, once the logical path to solve a problem is determined, current is passed through a circuit that remains static. During serial computation, logical operations are performed in a strictly defined sequence to obtain the solution. In contrast, circuits of biological processors are dynamic during computation and all fundamental computing elements operate collectively and simultaneously. As an alternative to serial logic operation, von Neumann demonstrated parallel computing on a piece of graph paper by moving black and white dots together using simple rules.[2-4] To implement such a cellular automaton (CA) in hardware, each cell representing the dots should communicate with its neighbors simultaneously to generate a collective decision.[5-6] Proposals are mounting for such network-based molecular computing.[7-9] Recently, suitable cells/dots have been built using Quantum dots and molecules.[10-13] However, thus far, it has not been possible to assemble them into a 2-dimensional grid where a CA cell communicates with many other devices at a time to execute a collective computation. In addition, in conventional cellular computing, the rules for updating cell states are fixed prior to computation, and the circuit or neighborhood is kept static. Realization of dynamic CA circuits would enable us to address problems that are prerogatives of natural bio-processors.

Here, instead of wiring single molecules/CA cells one-by-one, we directly build a molecular switch[14-17] assembly[18-23] where ~300 molecules continuously exchange information among themselves to generate the solution.[5] We demonstrate the ability of a CA cell to change its neighborhood from 2 to 6 CA cells in a controlled manner. Physically, it means that a molecule could interact locally with up to 6 molecules at a time during information processing. It should be noted that in the human brain, a neuron communicates with up to ~$10^4$ neighboring neurons at a time, and neural circuits evolve continuously during the life-span. By separating a monolayer from the metal ground with an additional monolayer, we developed a generalized approach to make the assembly sensitive to the encoded problem. The assembly adapts itself automatically for a new problem and redefines the CA rules in a unique way to generate the corresponding solution. We demonstrate the computing potential of the molecular assembly by realizing standard computational constructs like logic gates, directed propagation of information, etc. In addition, we have demonstrated the potential of dynamic circuit creation by physically encoding differential equations and generating quantitative solutions for heat diffusion[24-25] and mutation of normal cells into cancer cells.[26] The molecular assembly functions similarly to the graph paper of von Neumann, where excess electrons move like colored dots on the surface, driven by variation of free energy that leads to emergent computing.[27-30] The assembly processes ~300 dots at a time, whereas the fastest processors today operate only one bit/dot at a time per channel.

## Realization of a DDQ computing grid

The *2,3-dichloro-5,6-dicyano-p-benzoquinone* (DDQ) molecule (Fig. 1a) that reversibly switches among the four conducting states[17,18] 0, 1, 2 and 3 is deposited on an Au (111) surface as a bi-layer (Supporting Information, SI a). To encode information, the molecules are selected one-by-one by physically moving the *scanning tunneling microscope* (STM) tip to the highest current location on the molecule as shown in Fig. 1b. Then the desired conducting/logic states 0, 1, 2 or 3 are written by applying a pulse of -1.6, 1, 1.3 or 1.6 V tip bias respectively for 5-10 µs at ~300 K. State 3 appears as a brighter sphere in the STM image due to having an additional trapped electron than in state 1. State 2's brightness is between that of state 1s and state 3s as it has a more elongated prolate-shape than state 0. Fig. 1c shows detection of four states. The STM image is represented by a 2D map (matrix) of states 0, 1, 2, and 3, which are denoted by blue, green, yellow, and red balls respectively (Fig. 1d).

The connecting region between two neighboring DDQs with more than 60% of the peak tunneling current observed on the molecules is considered as a wire, which is represented by a solid line in Figs. 1e and 1f. Thus, we get a circuit where one molecule is connected to a distinct number of neighboring molecules with which it interacts at a time. This interaction is referred to as *one-to-many interaction at a time* in this article. The neighborhood of a molecule or CA cell varies from 2 to 6 in the eight molecular circuits possible in the top monolayer. Continuous scanning of the bi-layer by changing the tip bias from -2 V to +2 V in a loop reveals eight distinct circuits (Fig. 1g). In a matrix, if states 0 or 2 are in excess of 60% within ~20 nm$^2$ area, the DDQs prefer to reassemble into a circuit of Type 7 or 2 respectively. If the number of state 1s are more than 50% or the number of state 3s are more than 30 % in a ~20 nm$^2$ area, then a circuit of Type 1 or 5, respectively, is created (Figs. 1f, g). As we write a matrix, depending on the concentration of excess electrons (state 1s and 3s), DDQs in some part of the matrix region re-orient to the nearest energetically favorable circuit. Hence, for every new input matrix, a unique arrangement of several circuit-domains is created automatically (Fig. 1g, Movie 3). Thus the density of free electrons in an input pattern, the transformed circuit and the logic state transport rules are correlated.

Switching of an entire molecular arrangement as a function of applied bias has been reported for different molecules (SI b).[19-23] However, the co-existence of multiple circuits side-by-side, as demonstrated in Fig. 1g, has not been observed. Due to weak inter-layer coupling (Fig. 1b) and strong coupling among surface molecules, the top monolayer relaxes almost independently. This leads to the survival of multiple circuits. Therefore, excess electrons supplied to the assembly via states 1 and/or 3 find themselves in a potential surface of valleys and hills;[5] this triggers their spontaneous motion to minimize the free energy. Since the encoding process re-defines the mode of interaction of DDQs with their neighbors, the excess electrons do not move randomly, but rather they follow well-defined rules (SI c). In particular cases, the spontaneous evolution of the input matrix requires an additional trigger by scanning the surface at -0.98 V. All states of the entire logic pattern are erased to state 0 by scanning the surface at -1.6 V, and it is possible to use the same surface repeatedly for pattern evolution.

## Logic state transport rules

We have identified the following seven categories of rules (Fig. 2) that dictate logic-state transport. **Rule 1:** If negatively charged state 1s and 3s are distributed heterogeneously in an input pattern, a pseudo positive charge (PPC) is generated to which they are attracted from at most ~15 molecules apart. As a result, state 1s and 3s move to this single point. Since ~ 700 molecules influence each other at a time, it is not feasible to express this Rule in an abstract form; it would require ~$4^{700}$ rules. Therefore, we recourse to an analytic approach to program this Rule (SI h). The dynamic feature of the rule is provided in Movie 4. **Rule 2:** A state 2 site offers a lower barrier for the electron transport than a state 1 site, and a state 1 moves faster than a state 3. By tuning the distance Δ$x$ between two DDQ molecules (0.93 to 1.03 nm) in an assembly of circuits we can speed up or slow down the electron transport process as described in Fig. 2 (SI d). To program this rule in our simulator such that the temporal behavior of transition state dynamics is consistent with our experiment, we leave a fraction of states 1 and 3 unchanged in each update. **Rule 3:** State 1s and/or 3s and their groups experience repulsive forces in a particular direction if they come in contact at a single point. This rule promotes divergence of bits in a logic pattern. **Rule 4:** State 1s and/or 3s and their groups move leaving a trail of state 2s that remain static. If seven state 2s form a hexagonal cluster, then all switch to the lower energy state 0 in two steps except the central DDQ. **Rule 5:** In particular arrangements, clusters of state 1s and/or 3s form a group and move on the surface as a single entity. Groups lose their property of a single entity if the contact dimension is at least two-molecules in length. **Rule 6:** When negatively charged states form a cluster that is asymmetric, the cluster changes to a symmetric shape through the creation of two state 1s by breaking one state 3 or through the creation of a new state 3 by fusing two state 1s. **Rule 7:** Depending on the charge density of an input pattern, a unique composition of circuits is formed. In a CA grid, a particular set of rules is favored in a typical circuit-domain (see tables in Figs. 2 and 4). The order of relative circuit areas for an input pattern determines the order of execution of Rules. The algorithms used to program these rules in our simulator are described in the online text (SI h).

## Conventional CA computing

*Writing, erasing and retrieving information:* In Fig. 3a we demonstrate the sequential writing of a state 1 matrix on a state 0 surface. The states are stored as static information until spontaneous pattern evolution is triggered externally. By scanning the surface at -1.68 V one can reset all molecules to state 0, thus erasing the information. To retrieve information the surface is scanned at ~0.2 V (Figure 1d).

***Directed propagation of Information:*** To send a complex information packet in a particular direction, we need to write additional states 1s and/or 3s so that an electron density gradient is created along the desired direction of propagation (Figure 3b top). One example of transport over an apparently unbounded distance is shown via simulation and experiment in Figure 3b (bottom). However, it is possible to send a packet to a particular location. The signal propagation stops, when the motion of states within the packet generates a pattern that periodically repeats in a fixed space. To send an information packet that is in the form of a group, we need to create its mirror image in contact with it; repulsion would then triggers the propagation of information following Rule 3, which dominates in the circuits 1, 5 and 7 (Figure 4a).

***Logic gate:*** The effects resulting from Rule 3 appear similar to the interactions in the Billiard Ball Model,[31] which has been used to design logic gates. We have realized an AND logic gate based on interactions in which Rule 1 dominates over Rule 3. A schematic logic device is shown in Figure 3c. A random composition of states 1 and 3 (density > 0.5 electrons/$nm^2$) written in a circular form is a logical "1" and the absence of any such composition is a logical "0". If we write two logical "1"s at most 15 nm (15 cells) apart, only then a new composition is created automatically on the surface depicting logical state "1". If any one or both input compositions are absent we do not get such an output; rather states collapse at the same space (Figure 3d), we get a logical "0" at the output location. Thus we realize an AND gate. A large number of such logic gates could be operated in parallel on DDQ CA by separating those by ~15 CA cells. The output of a logic gate could be transported to the input of another logic gate, as described in Figure 3b.

***Density Classification:*** Density classification (DC) is a task to categorize particular elements in a mixture. DC is challenging when elements of different densities interact and influence each other at the same time. Classically, in CA, minority states in a space are converted into the majority's state to reflect classification.[32] In our CA, however, we divide local space into multiple strictly defined regions on the basis of charge density, as shown in the table of Figure 4a. We can categorize the region into 8 circuit classes with a maximum sensitivity of 1 electron/20 $nm^2$ (Figure 4b). We have converted a state 0 space into two distinct circuit patterns simply by writing two input patterns with distinct charge densities. The CA grid spontaneously classifies two distinct density distributions, difference of ~2 electrons/30 nm$^2$, by converting the grid into two distinct circuit domains (Figure 4c).

***Voronoi Decomposition:*** In Voronoi decomposition[33] a space is divided into multiple regions; each of which corresponds to a point such that any two neighboring points are equally separated from the boundary between them. As described earlier, an organic monolayer when imaged at a low bias (~0.2 V) reveals the composition of circuits generated by the input pattern. The borders between the circuit domains are remarkably linear (see Figure 1e and Figure 4d). The Density Classification at larger scales (~50 nm$^2$) triggers segmentation of the computing space and decomposes the surface into Voronoi cells. A physical process in the molecular assembly could be modeled in terms of the evolution of the points (black dots in Figure 4d) representing the Voronoi cells.

## Mimicking two distinct natural phenomena

Biological computers like our brain do not have logic gates yet they solve complex problems. During computation, the encoded information pattern in the neurons dynamically modulates the neural architecture and continuously evolves to reach a collective solution. Theoretically, by tuning input patterns and CA rules, we can solve several problems without using any logic gates.[34] Here we mimic two natural events in the molecular CA matrix by tuning input patterns and effective CA rules. We encode two distinct input patterns that evolve over time in such a way that the transport of free electrons (or logic states) follows the essential features of diffusion[24-25] for one input pattern and follows the evolution of cancer cells[26] for the other input pattern. We have also emulated the pattern evolution in a simulator. The global features of the experimental patterns are in reasonably good agreement with the simulated patterns (SI e).

In the diffusion process, a blue ball (state 0) denotes a normal material. When it accepts one electron, it turns to green (state 1). Two excess electrons turn blue to red (state 3). When electrons leave, green and red balls relax to state 2s, which are yellow (Fig. 5a). To create a directional flow, we write a pattern of straight-lines each composed of alternating state 3 and state 1 so that they form a group and move as a single unit (Fig. 5b). The direction of flow is controlled by the potential gradient of states in the background.[24,25] For free energy minimization, the linear arrangement tries to bend into a circular shape, but as it begins to move, the coupling breaks. The broken parts follow the potential gradient independently (Movie 5). We calculate the concentration variation and the rate of electron flow as described in the Method Section and plot them in Fig. 5a. A linear relation suggests a diffusion process; its slope $D$ (2 nm$^2$/min) is the electron diffusion coefficient. Fig. 5b shows that gradually over time states 1 and 3 spread homogeneously on the surface. From Fig. 5c we get flux $\varphi(x,t) = \frac{14}{t^{1/2}} \exp(-\frac{(x-5.5)^2}{t})$ as the quantitative formal solution of the diffusion equation on a linear array of 10 cells.

In a conventional CA, since transition Rules are fixed, irrespective of the input pattern one can generate only one kind of dynamics of logic states.[3-5] To demonstrate the potential of CA rule modulation during computation (Rule 7, Fig 2), we generate a significantly different kind of dynamics in the same CA grid by emulating the mutation of normal cells to cancer cells. To formulate the mutation process, we correlate four transition states of the normal cell to four conducting states of a DDQ. A normal cell (state 0) first mutates inactivating a single *tumor suppressor gene* (TSG) (state 1) at a rate $u_1$; it then mutates again inactivating another TSG (state 3) at a different rate, $u_2$ (Movie 5, Fig 6a). Cellular proliferation occurs when one state 3 creates two cells in state 1. Depending on the input pattern, the fusion of two state 1s into one state 3 may also dominate. When cancer spreads in tissues, it may leave a trace of chromosomal instability (CIN) or state 2. In theory, the inactivation of TSGs follows three successive differential equations.[26] The solution for the third equation (see Methods section) provides the number of cancer cells (state 3s) produced, which in our CA grid is the count of new red balls generated (N3). N3 depends on the cell population, $N$, in a tissue compartment (Fig. 6b). We write state 1s in the form of two concentric rings so that they neither converge nor diverge (Fig. 6c). Thus, an artificial tissue boundary CG is established, wherein $N$ is the total number of DDQs inside CG. We tune $u_1$ by switching some of the state 0s to state 1s inside CG between every two scans, whereas $u_2$ is decided by the system itself.

We change $N$ by modifying the separation between the two rings while keeping the inner ring unchanged. Here, the half-life $t_{1/2}$ for state 3 is the duration for which N3 is increased to 2N3. We plot average $t_{1/2}$ values for different values of $N$ in Fig. 6a. A similar feature between half-life $t_{1/2}$ vs. $N$ plot in Fig. 6a and the kinetics of cancer[26] suggests a consistent encoding of a two rate controlled phenomenon in the CA grid.

Fig. 6c shows that, at a very low population ($N < 290$), when the two rings are less than 3 nm apart, the inner ring stabilizes by rearranging the states after which both the rings merge. In this case, almost every collision between two state 1s produces a state 3, and N3 varies with time as $\sim 4t^2$ (Fig. 6b). The conversion continues at this rate until all state 1s are converted to state 3s. For an intermediate population ($290 < N < 620$), the inter-cluster distance increases to ~5 nm, and the state 1s of two rings collide before reaching an equilibrium. Instantly produced state 3s break into state 1s, and N3 varies as $\sim 18t$. For a very high population ($N > 620$), when two rings are more than 8 nm apart, a higher $PPC$ leads to a large-scale one-to-one collision of state 1s; thus conversion to state 3 is sped up. Therefore, the abundance of cells in state 1 allows N3 to vary as $\sim 40t^2$. N3s in Fig. 6b provide quantitative solutions for the differential equations. During evolution, if state 2s are erased or written, the production of state 3s is also slowed down or sped up respectively; this is a distinct feature of CIN dynamics. As N3 kinetics changes with time, different sections of the lines in Fig. 6b follow additional solution functions 7e-4$t^2$, 7e-10$t^3$…. $t^5$ for low populations and 0.05$t^2$, $t^2$, $t^{1.8}$…$t^{1.2}$ for high populations. Remarkably, these smaller parts reveal complex CIN dynamics, which are consistent with the established models.[26]

During realization of diffusion and cancer growth on a 24×27 molecular-matrix, changes in the STM image contrast at 300 DDQ sites reveal that the Rules are executed at least at 300 sites simultaneously.[18] Cancer may evolve naturally in real tissue of ~ $10^9$ cells in ~ 100 years, and heat diffuses in metals at a speed of ~ 100 km/s. However, in a molecular assembly, ~600 cells mutate in 6 minutes, and heat diffuses ~10 nm in 5 minutes. By writing an input matrix, we can tune the grids $\Delta x$ and hence $\Delta t$, and solve several differential equations similar to conventional methods.

## Outlook

In order to realize a CA that can carry out a wide variety of computational tasks, it is necessary to obtain a sufficient level of control on the transition state dynamics. The DDQ CA in this article provides an intriguing avenue to achieve such control while still relying on a relatively small and simple molecule. It is not only the interactions of the molecules in the DDQ CA but also the subtle formation of circuits in the molecular top-layer that count in facilitating transitions between states. The evolution of circuits is dependent on an easy-to-control parameter: the charge density in an area. These circuits dramatically influence the dominance of transition rules, as we have observed, and offer an efficient way to influence the computational behavior of the CA. The robust functioning of local circuits originates from the CA cell's *one-to-many communication and interaction at a time.*[30,18] Generalization of this principle would change the existing concept of static circuit-based electronics and open up a new vista of emergent computing using an assembly of molecules.[27-29,6]

**Methods:**

**Correlating Diffusion parameters with the assembly states:**

The rate of change of electron flux $\varphi$ is proportional to the variation of the gradient of electron flux, or $\frac{\partial \varphi}{\partial t} = D\nabla^2 \varphi$; $D$ is the diffusion coefficient. Here, $\nabla^2 \varphi \sim \frac{\partial^2 \varphi}{\partial A^2}$ where $A$ (~ 2.4 nm$^2$) is the area enclosed by seven DDQ molecules (unit cell, Fig. 5a).

In a unit cell, $\varphi$ is the ratio of total excess electrons $Q$ ( = *number of State*1 + 2 × *number of State*3 ) and $A$. Now, $\frac{\partial \varphi}{\partial A} \approx \frac{\Delta \varphi}{UC}$; $\Delta \varphi$ is $\Delta Q$ between two adjacent unit cells say 1 and 2 (or 3 and 4) of Fig. 5a (inset - bottom), and $UC$ is the non-overlapped region of two unit cells (~1.3$A$). The difference in $\frac{\partial \varphi}{\partial A}$ between two neighboring unit cells say 1 and 3 of Fig. 5a (inset) is calculated as $\frac{\partial^2 \varphi}{\partial A^2}$ and the value is assigned to the center of unit cell 2 where $\frac{\partial \varphi}{\partial t}$ is measured. Here, $\frac{\partial \varphi}{\partial t} \approx \frac{\Delta \varphi}{\Delta t}$; $\Delta t$ (40 seconds) is the time lapse between two STM scans.

Similarly, for every unit cell of the 24×27 matrix, we calculate $\frac{\partial^2 \varphi}{\partial A^2}$ from an STM image at $t = t_0$ and $\frac{\partial \varphi}{\partial t}$ from two STM images at $t = t_0$ and $t = t_0 + \Delta t$. For each value of $\frac{\partial^2 \varphi}{\partial A^2}$, the corresponding values of $\frac{\partial \varphi}{\partial t}$ are plotted in Fig. 5a. We have also plotted $\varphi$ vs. ranking Z (0, 1, 2….n) of the central molecule of the unit cells along a straight line in Fig. 5c. Therein, electron flux diffusion saturates in 4 minutes 40 seconds (noted as 4.40). We have taken the weighted average of flux to convert the staircase characteristic into a continuous trace in Fig. 5c; this is essential to find a formal quantitative solution. Equations of higher order and/or degree can be solved similarly.

**Correlating Cancer cell Evolution parameters with the assembly features:**

We have three coupled differential equations and two variables: states 1 and 3. Since state 1s can not be generated spontaneously, we encode state 1s using STM between two scans at a rate of 1.75×10$^{-2}$ minute$^{-1}$ so that a mutation rate of $u_1$ (1.3×10$^{-1}$ minute$^{-1}$) for state 0 is maintained. The evolution of state 0 follows $\frac{dX_0}{dt} = -u_1 X_0$, where $X_0(t)$ is the number of effective state 0s at time $t$, given by $X_0(t) = X_0(0)\exp(-u_1 t)$. Here, $X_0(t=0) \approx$ (initial state 1s in the rings + externally encoded state 1s)/2, which is the number of effective normal cells (state 0s) involved in generating state 3s. The state 1s that leave the rings and the state 1s encoded via STM to compensate for the loss during state 3 production contribute together to the evolution of state 1, which is described by $\frac{dX_1}{dt} = u_1 X_0 - N_{eff} u_2 X_1$ ; its solution is $X_1(t) = X_0(0)[u_1(e^{-u_1 t} - e^{-N_{eff} u_2 t})/(N_{eff} u_2 - u_1)]$, where $N_{eff} = 2X_0(t=0)$. State 1 mutates at a rate $u_2$ to produce $X_3$ number of state 3s following $\frac{dX_3}{dt} = N_{eff} u_2 X_1$, which yields N3= $X_3(t) = X_0(0)[1 - (N_{eff} u_2 e^{-u_1 t} - u_1 e^{-N_{eff} u_2 t})/(N_{eff} u_2 - u_1)]$. If state 2s are deleted between two scans, $u_2$ decreases to ~$1.18 \times 10^{-5}$ minute$^{-1}$ from ~1.79, ~2.70, and ~3.12 ($\times 10^{-3}$) minute$^{-1}$ for $X_0(0) = 128$, 144, and 157 respectively. Thus, CIN increases $u_2$ by ~$10^2$ times, which prompts us to simplify $X_3(t)$. For a smaller time scale that is considered in our experiment, a small population yields N3~ $X_0(0) N_{eff} u_1 u_2 t^2 / 2$ ; an intermediate population yields N3 ~ $X_0(0) N_{eff} u_1 \sqrt{u_2} t$ ; and a large population yields N3~ $X_0(0) N_{eff} u_1 u_2 t^2 / 2$.[26] The number of new red balls, N3, is obtained by comparing two consecutive STM images. The average values of N3 obtained by repeatedly evolving the same input patterns as demonstrated in Fig. 6c are plotted in Figs. 6a, 6b. In Figs. 6a and 6b, $N$ = (state 1s + state 0s) initially inside the CG.

**End Notes:**

**Acknowledgement:** Authors acknowledge H. Hossainkhani, Y. Wakayama, J. Rampe, M. McClain, W. Cantrel and J. Liebescheutz for discussion. The work is partially funded by the Ministry of Science and education, culture and sports (MEXT), Japan during 2005 (Feb) -2008 (Mar) and Grants in Aid for Young Scientists (A) for 2009-2011, Grant number 21681015. R.P. acknowledges National Science Foundation (NSF) Award number ECCS-0643420.

**Declaration:** There is no competing financial interest among the authors.

**Contribution:** A.B designed research; A.B did the experiment; A.B developed CA simulator; R.P, A.B and S.S did the theoretical studies; A.B, R.P, S.S and F.P analyzed the data; and A.B, R.P, F.P and S.S wrote the paper together; D.F reviewed the work.

**Correspondence**: *To whom correspondence should be addressed. E-mail address: anirban.bandyo@gmail.com and anirban.bandyopadhyay@nims.go.jp

**Supporting Online Material**

Supporting online materials are available for this article in the website link.

Supporting online text file contains additional information as referred SI in this manuscript. If there is any problem in viewing the movie files, please download codec from www.divx.com/codec

Movie 1. *Formation of bilayer*

Movie 2. *Creation of Matrices*

Movie 3. *Construction and re-construction of circuits*

Movie 4. *Cellular Automaton Rules*

Movie 5. *Diffusion and Evolution of cancer*

**Figure Captions:**

**Figure 1. The concept of a wireless molecular circuit:**
**a**. The DDQ molecule. **b**. The DDQ bilayer's atomic structure; side view (above); top view (below, Movie 1). T denotes a molecule on the top layer. **c**. STM images for four states (top); corresponding ball models and atomic structure (bottom). Current profiles of four states (black line) taken along the white lines having current heights of (at 0.68 V) ~ 0.15 nA (state 0), 0.25 nA (state 1), 0.28 nA (state 2) and 0.32 nA (state 3) respectively. The height profiles are ~ 0.9 nm wide. **d.** A 13×17 matrix (~400 nm$^2$) and its ball representation (right). **e**. STM image at 0.2 V tip bias and 0.05 nA tip current. **f**. Zoomed region of the STM image and the corresponding circuit (below). The black rings represent molecules; number represents interconnecting lines (wires). **g**. Eight possible circuits (left) in image **e** are detected (right). The red lines are Voronoi cell boundaries.

**Figure 2. Discrete logic-state transport rules:**
**Rule 1: Convergent Universe**: A charge moves a distance *d* towards *PPC* (left). To calculate *PPC*, we neglect those areas that have charge(*s*) in between (right). **Rule 2: Creation of spatial Δx and temporal limit Δt:** The <Δx> (experiment) and <Δt> (simulation) are plotted for four surfaces initially covered with state 0s, 1s, 2s, or 3s. **Rule 3: Divergent Universe:** Examples of collisions; arrows denote the direction of motion of logic-states. **Rule 4: Life of logic states:** Trail of state 2 for motion of state 1 and 3 (top); the death of a cluster of state 2 (bottom). **Rule 5: Collapse of space:** Four examples of group formation of state 1s, 3s. **Rule 6: Transformation:** Fusion of two state 1s to create a state 3 (left); breaking of state 3 to create two state 1s (right). **Rule 7: Priority of Rules:** Correlating input patterns, initial circuits and the dominant Rules (see Movie 4 for details and SI text online for the algorithm to program these rules).

**Figure 3. Information encoding, retrieving, transport and Logic Gate operation:**
**a.** STM image (left) and its equivalent ball representation (right) are shown in four steps (top to bottom) of an event where all state 0s in a 6×7 grid are converted into state 1s. **b**. Top row - schematic representation of an information packet. The shaded region has a higher electron density, which determines the direction of motion. A real packet is shown to its right. Middle row- a simulated transport of an information packet comprised of a few state 1s (left-input).

Yellow balls trace the transport path for 2000 steps (right-output). Bottom row- STM images for initial (left) and final steps (right) of information transport (scale bar 7.5 nm). **c**. Truth table for an AND gate (left). STM images (right; scale bar 2.8 nm) show A=1, B=1, C=0 (top-input) and A=1, B=1, C=1 (bottom-output). **d**. AND gate operation with [A=1, B=0] and [A=0, B=1] should not have evolved state at location C; A or B should retain logic state 1. 35 state 1s used to create A; simulation shows that state 1s collapse to maintain the logic state 1 (A=1).

**Figure 4. Computing constructs, Density classification and Voronoi Decomposition:**
**a.** Relations among density distribution, circuit type and dominating CA rules. **b**. Schematic representation: all CA cells in state 0 (circuit 7, left); the encoded input pattern (middle); the derived composition of circuits (right). **c**. STM image of a CA grid (6×5) comprised of state 0s (circuit 7), ball representation below. Two patterns (colored balls, right) are encoded. One pattern converts the grid to circuit 1 (top) another to circuit 4 (bottom). Scale bar is 1 nm. **d**. 13×15 CA grid at state 0 (left); a pattern is encoded that converts it into a composition of circuits (middle) noted by circuit number. Voronoi decomposed cells containing the Voronoi points (black dots) derived from the middle image (right).

**Figure 5. Mimicking natural phenomenon I: Electron diffusion:**
**a.** Electron flux rate $\frac{\partial \varphi(t)}{\partial t}$ vs. surface-electron concentration variation $\frac{\partial^2 \varphi(t)}{\partial A^2}$. Inset (top) - transition scheme of DDQ logic-states. Inset (bottom) - a linear array of unit cells and their ranking (1, 2, 3, 4). The shaded green and blue regions are unit cells. **b.** STM images (24×27 nm) at 0.68 V tip bias and 0.05 nA tip current (right) with corresponding simulation (left) shown side-by-side with time (minute) (Movie 5). **c.** Temporal (minute) variation of electron concentration $\varphi$ along a linear array of 10 unit cells (Z: 1 to 10) in the white ring of panel **b** is plotted.

**Figure 6. Mimicking natural phenomenon II: Evolution of cancer cells:**
**a**. Measured half-life $t_{1/2}$ (seconds) to produce state 3 vs. tissue population *N* on a log-log scale. First row (inset) - a real cell with green lines as TSG with corresponding ball representation (second row). Third row – a scheme for 2 *one-TSG-deactivated cells creating one cancer cell and one CIN cell*. **b**. Number of State 3s produced (N3) vs. time for different *N*. **c**. Simulated (left) and corresponding STM images (24×27 nm) at 0.68 V tip bias and 0.05 nA tip current (right) are paired and ordered in a row with time (minutes). Simulations with $N$ = 286, 456, or 627 having 149 (~50%, top), 196 (~40%, middle), or 222 (~34%, bottom) state 1s respectively. After each STM scan of 40 seconds, 5 (top), 8 (middle), and 11 (bottom) new state 1s are written inside the respective CGs within 20 seconds (Movie 5).

Figure 1

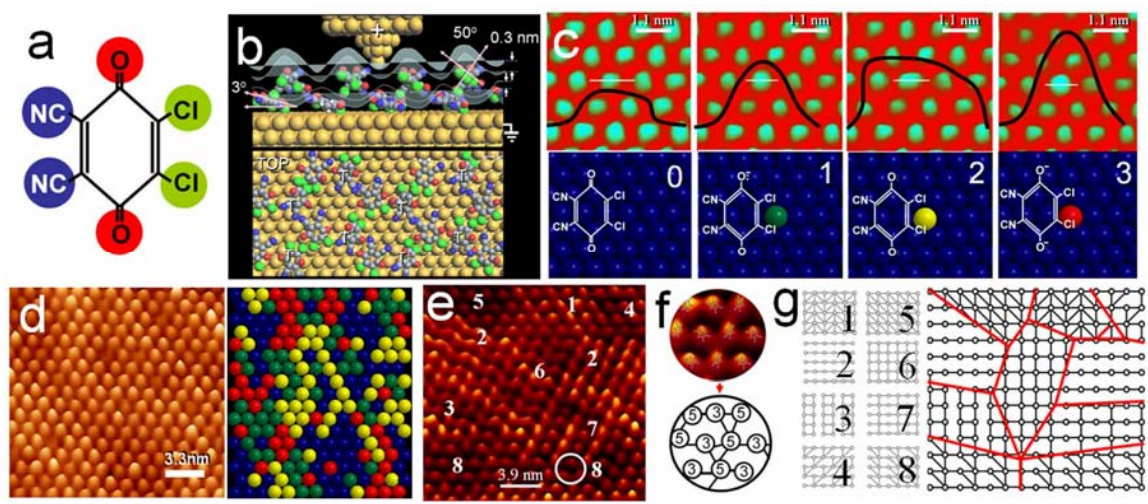

Figure 2

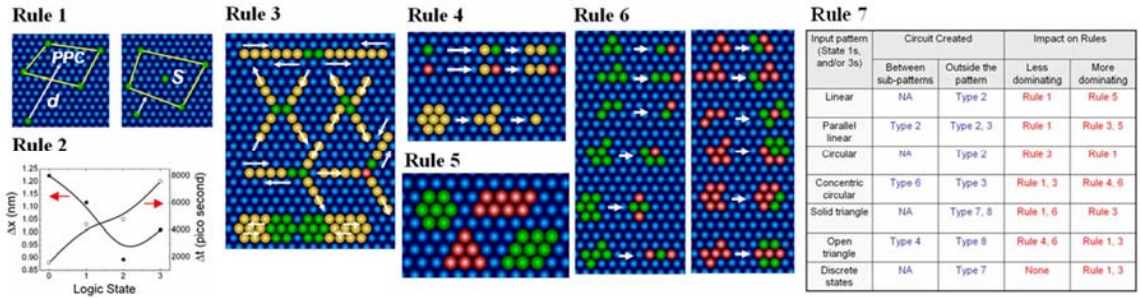

Figure 3

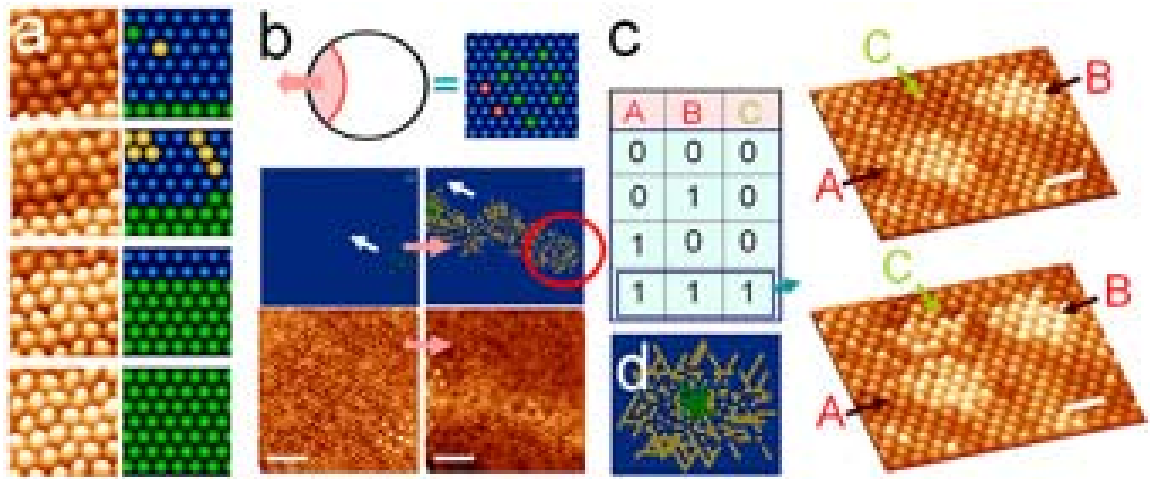

Figure 4

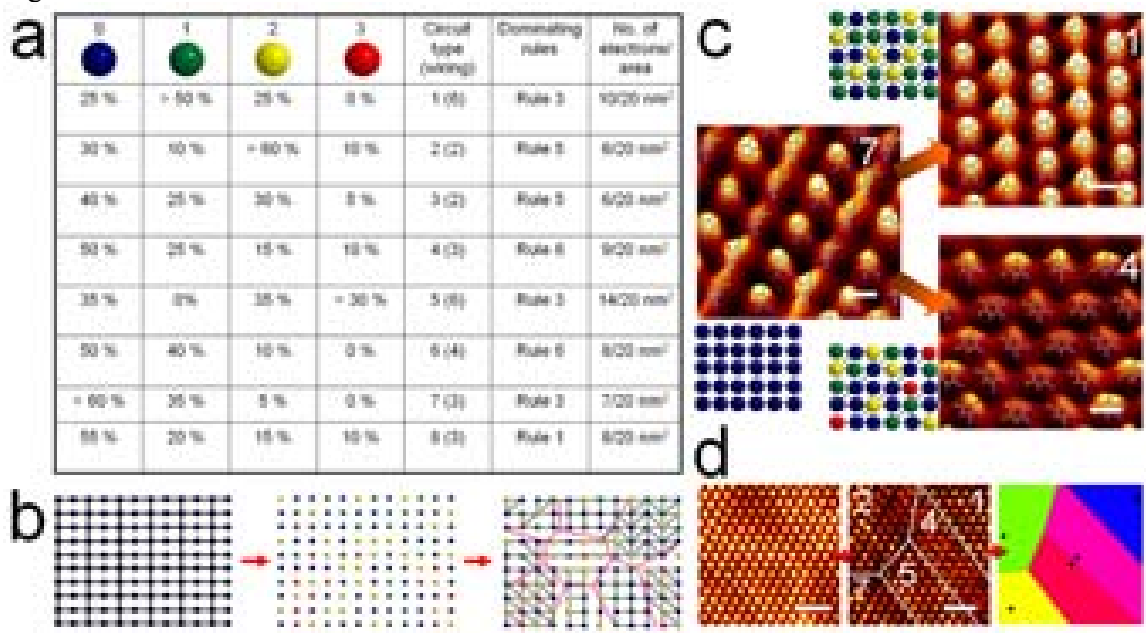

Figure 5

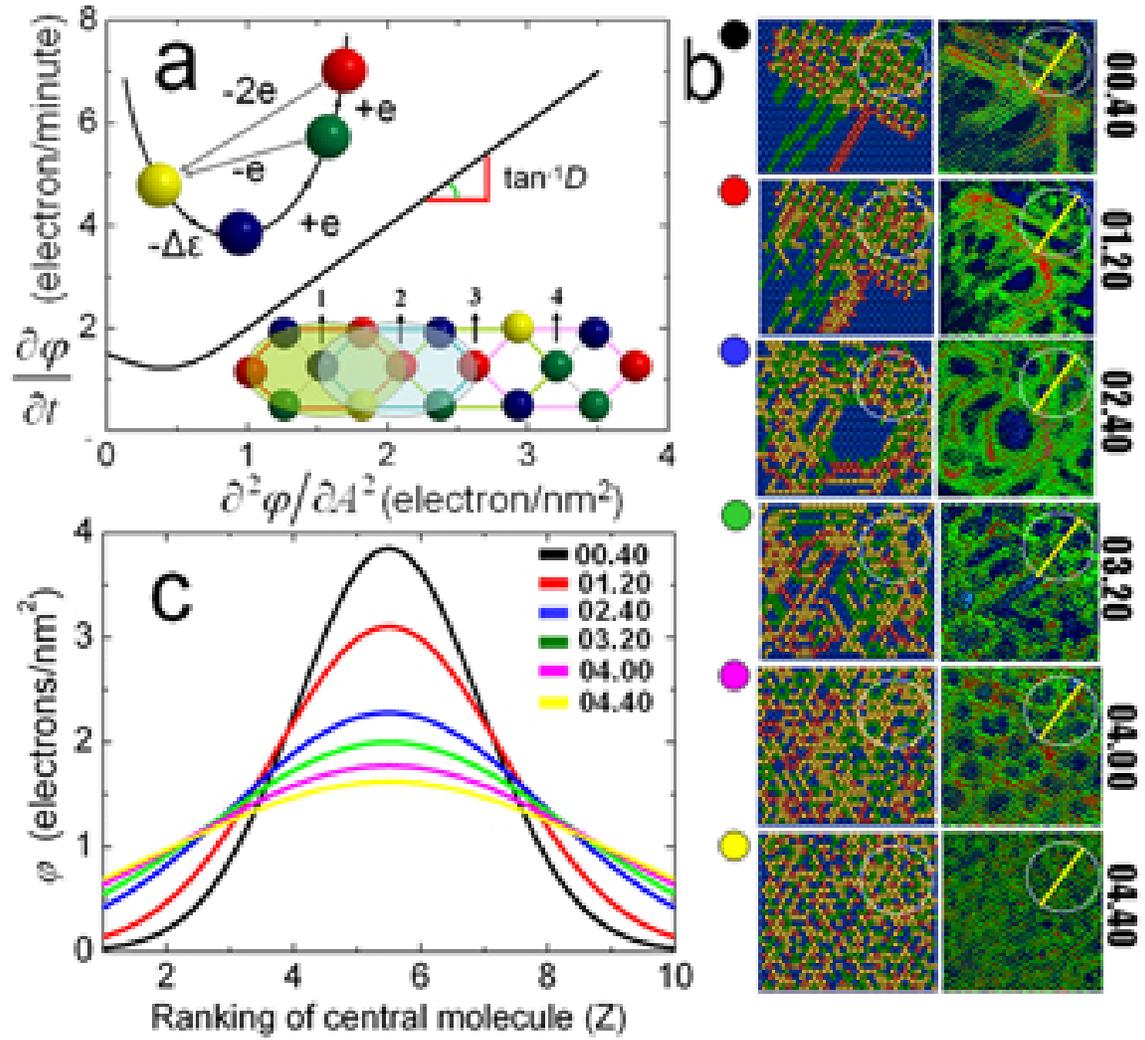

Figure 6.

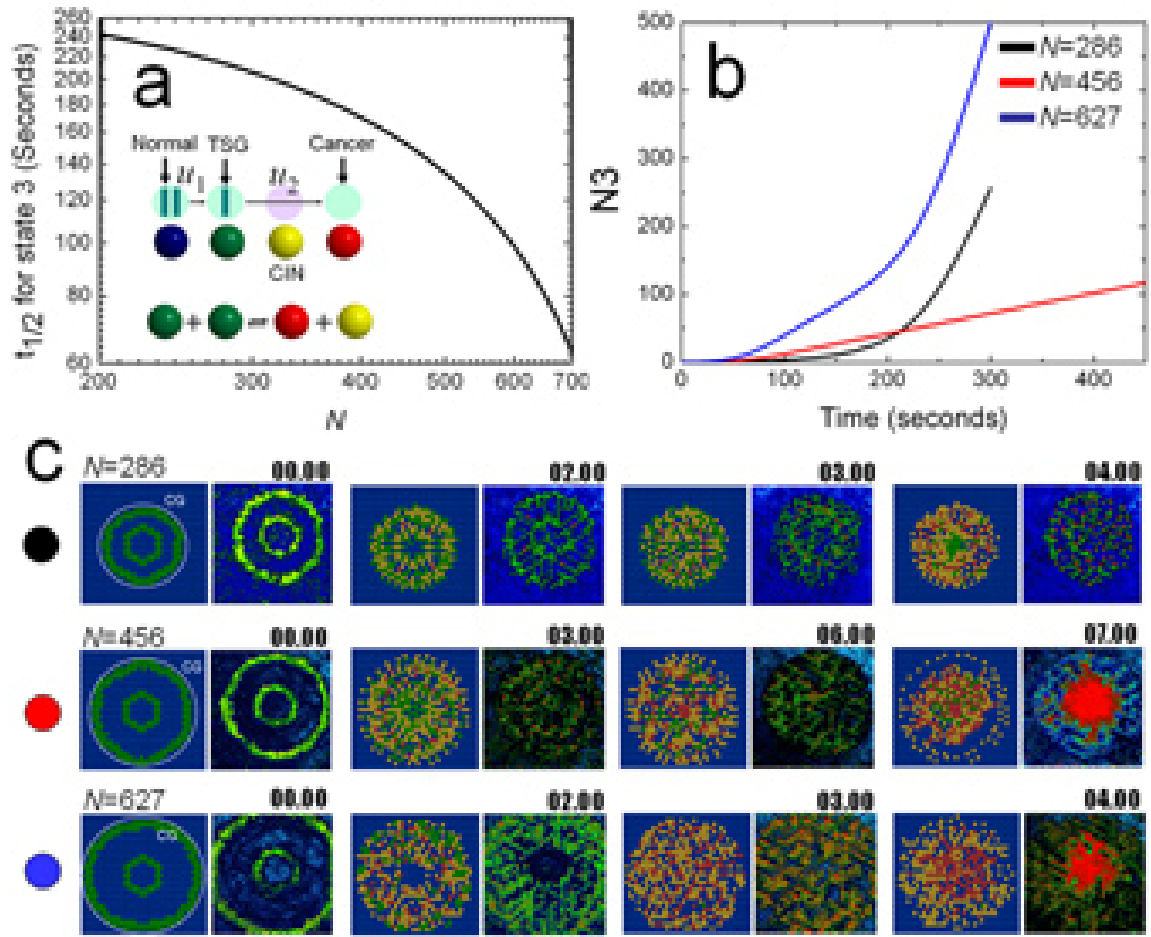